\documentclass[12pt]{article}
\usepackage{}
\usepackage{graphicx}
\usepackage{amsfonts}
\usepackage{amssymb}
\usepackage{mathrsfs}
\usepackage{amsmath}

\usepackage[unicode]{hyperref}
\usepackage[numbers,sort&compress]{natbib}

\begin{document}
\renewcommand{\thefootnote}{\fnsymbol{footnote}}
\begin{titlepage}

\vspace{10mm}
\begin{center}
{\Large\bf Complexity/Action duality of shock wave geometry in a massive gravity theory}
\vspace{8mm}

{{\large Yan-Gang Miao${}^{1,2,3,}$\footnote{\em E-mail address: miaoyg@nankai.edu.cn}
and Long Zhao}${}^{1,}$\footnote{\em E-mail address: longzhao@mail.nankai.edu.cn}

\vspace{6mm}
${}^{1}${\normalsize \em School of Physics, Nankai University, Tianjin 300071, China}

\vspace{3mm}
${}^{2}${\normalsize \em Frankfurt Institute for Advanced Studies (FIAS),\\
Ruth-Moufang-Str. 1, 60438 Frankfurt am Main, Germany}

\vspace{3mm}
${}^{3}${\normalsize \em Institut f\"ur Theoretische Physik, Goethe-Universit\"at Frankfurt am Main,\\
Max-von-Laue-Str. 1, 60438 Frankfurt am Main, Germany}
}

\end{center}

\vspace{4mm}
\centerline{{\bf{Abstract}}}
\vspace{4mm}
On the holographic complexity dual to the bulk action, we investigate the action growth for a shock wave geometry in a massive gravity theory within the Wheeler-De Witt (WDW) patch at the late time limit. For a global shock wave, the graviton mass does not affect the action growth in the bulk, i.e. the complexity on the boundary, showing that the action growth (complexity) is the same for both the Einstein gravity and the massive gravity. Nevertheless, for a local shock wave that depends on transverse coordinates, the action growth (complexity) is proportional to the butterfly velocity for the two gravity theories, but the butterfly velocity of the massive gravity theory is smaller than that of the Einstein gravity theory, indicating that the action growth (complexity) of the massive gravity is depressed by the graviton mass. In addition, we extend the black hole thermodynamics of the massive gravity and obtain the right Smarr formula.


\vskip 10pt
\noindent
{\bf PACS Number(s)}: 04.50.Kd, 04.60.Bc, 04.70.Dy


\end{titlepage}
\newpage
\renewcommand{\thefootnote}{\arabic{footnote}}
\setcounter{footnote}{0}
\setcounter{page}{2}

\section{Introduction}

The holographic principle shows~\cite{9711200} that the bulk dynamical
evolution can be coded in the boundary field theory without gravity.  The black hole interior evolution is related to the boundary transverse entanglement that reaches to its maximum value at the scrambling time~\cite{1306.0622}. Based on the characteristic that the interior volume of black holes grows linearly with respect to time, Susskind then pointed out  \cite{1403.5695,1507.02287} that the black hole interior volume should be dual to the complexity of the boundary system, i.e. the Complexity/Volume (C/V) duality. If  the bulk spacetime contains a shock wave, the interior volume decreases in a specific period of time, and the observer falling into the horizon will collide the shock wave when he arrives at the horizon. So the complexity can be regarded as a criterion of existence of firewalls \cite{1406.2678}.

Recently, a new conjecture was proposed by Susskind and his collaborators~\cite{1509.07876,1512.04993}, in which the boundary complexity is connected to the classical bulk action in the Wheeler-De Witt (WDW) patch. The new assumption, referred to as the Complexity/Action (C/A) duality, can be expressed  as follows,
\begin{align}
C=\frac{A}{\pi\hbar},
\label{bound}
\end{align}
where $C$ is the boundary complexity\footnote{It implies the minimum numbers of quantum gates that are required to produce the corresponding state associated with such boundary complexity from the reference state.} in quantum information theory and $A$ is the total classical gravitational action in the bulk region within the WDW patch. Compared with the C/V duality, the C/A duality does not depend on any length scale chosen by hand, such as the AdS curvature radius $l_{\rm AdS}$ or the black hole horizon radius $r_{\rm h}$.

Given the energy of a quantum system, as already shown by Lloyd \cite{Lloyd}, the growth rate of the bulk action or the
computational rate of the boundary state should have an upper bound,
\begin{align}
{\rm the} \;{\rm computational} \; {\rm rate} \leq\frac{2E}{\pi \hbar},
\label{ca}
\end{align}
where $E$ is the excited energy of the boundary state. Substituting eq.~($\ref{bound}$) into eq.~($\ref{ca}$), one can obtain,
\begin{align}
\frac{dA}{dt}\leq 2E.
\label{bound of BH}
\end{align}
In the Einstein gravity, for instance, in the cases of the static spherical shell and the shock wave geometry, the computational rate has been checked~\cite{1512.04993}, where the equality meets for a neutral black hole.
This inequality has also been examined by Cai et al.~\cite{1606.08307}, where the universal holographic upper bound can be expressed by the difference between the value of thermodynamics quantities at the outer horizon and that at the inner horizon.

In the recent studies~\cite{1610.08063,1612.00433,1702.06766,1702.06796,1703.10468,1703.06297} on the holographic complexity dual to the bulk action in different gravity theories, the spacetime with a shock wave reflects more characteristics of the boundary complexity, such as the criterion of existence of firewalls.
On the other hand, when a boundary disturbance (perturbation)
depends on transverse coordinates, the precursor operator 
grows in spatial directions and the complexity caused by this disturbance is closely related to the growth velocity of disturbance in spatial directions. The growth velocity of disturbance is the so-called butterfly velocity. The butterfly velocity has been investigated by the calculation of the out-of-time order four-point function~\cite{1602.08272} in the cases of the Topologically Massive Gravity (TMG) and the New Massive Gravity (NMG)~\cite{1610.02890} in which the graviton contains a massive mode. In the TMG and NMG, the butterfly velocity $v_{\rm B}$ depends on the scaling dimension $\Delta$ in the following way,
\begin{align}
v_{\rm B}=\frac{s-1}{\Delta-1},
\end{align}
where $s$ equals two for the spin of graviton, $\Delta \equiv 1+\sqrt{1+l_{\rm AdS}^2m^2}$, and $m$ is the graviton mass. One can see that the butterfly velocity of a massive mode is smaller than that of a massless mode.

The above observations motivate us to examine the C/A duality of the shock wave geometry in massive gravity, such as whether the C/A duality associated with a shock wave is a general principle beyond the Einstein gravity theory, and also to study the effect of  graviton mass. We shall investigated the action growth in the bulk, i.e. the complexity on the boundary, for the shock wave geometry in the massive gravity~\cite{1011.1232} which contains only a massive mode.\footnote{The TMG and NMG contain both a massive mode and a massless mode. Under some restriction, the two gravity theories have only a massless mode.}
We find that the action growth (complexity) of the massive gravity in the case of the global shock wave is equal to that of the Einstein gravity
because the effect of the global shock wave shifts the Kruskal coordinate $v$ only a transverse-coordinate-independent quantity which does not depend on the graviton mass.
We also discover that the action growth (complexity) of the massive gravity in the case of the local shock wave is depressed by the graviton mass because the action growth (complexity) of the two gravity theories is proportional to the butterfly velocity while the butterfly velocity of the massive gravity is smaller than that of the Einstein gravity.
In addition, we extend the black hole thermodynamics of the massive gravity and obtain the right Smarr formula by using the C/A duality and the new calculation method of boundary terms~\cite{1610.02038}.

The paper is organized as follows. In section \ref{sec:shock}, we briefly review the black hole solution and the shock wave geometry for the massive gravity model in the 4-dimensional spacetime. In section \ref{sec:growth}, we give the action growth within the WDW patch in the late time limit. Then we extend the black hole thermodynamics of the massive gravity and give the right Smarr formula in section \ref{sec:volume}. Finally, we give a brief summary in section \ref{sec:conclusion}.

\section{Shock wave geometry in massive gravity}
\label{sec:shock}
Before calculating the action growth in the WDW patch, one should get the solution of black holes in a massive gravity theory. The action of the massive gravity~\cite{1011.1232,1301.0537,1409.2369,1511.04967} contains the Einstein-Hilbert action with a negative cosmological constant, the graviton mass terms, the Maxwell electromagnetic action, and the York-Gibbons-Hawking surface term, appearing in order as follows,
\begin{align}
S=\int d^4 x\left[\frac{1}{16\pi G}(R-2\Lambda)+\frac{m^2}{8\pi G}(\alpha_1 u_1 + \alpha_2u_2)-\frac{1}{16\pi}F^2\right]+\frac{1}{8\pi G}\int d^3 x\sqrt{-\gamma}K,
\end{align}
where $m$ is the graviton mass, $\gamma$ the induced metric on the boundary, and $K$ the trace of the extrinsic curvature. Note that $u_1$ and $u_2$ are associated with the graviton mass terms and can be expressed as
\begin{align}
u_1 &=tr \mathcal {K},\nonumber\\
u_2 &=\left(tr \mathcal {K}\right)^2-tr\left(\mathcal {K}^2\right),
\end{align}
where the matrix $\mathcal {K}$ is defined by ${\mathcal{K}^\mu}_\nu\equiv \sqrt{g^{\mu\alpha} f_{\nu\alpha}}$, and $f_{\mu\nu}$ is the non-dynamical reference metric chosen~\cite{1409.2369} to be $f_{\mu\nu}={\rm diag}\left(0,0,1,\sin^2\theta\right)$. The graviton mass terms destroy the differemophism invariance in the transverse directions ($\theta$, $\phi$) of spherical coordinates but keep the invariance in $t$ and $r$ directions. Parameters $\alpha_1$ and $\alpha_2$ are chosen \cite{1409.2369} to be negative in order to guarantee the existence of the Hawking-Page phase transition and of the extremal configuration of the black hole with zero temperature. The equations of motion derived from the above action read
\begin{align}
R_{\mu\nu}-\frac{1}{2}g_{\mu\nu}R-\frac{3}{l_{\rm AdS}^2}g_{\mu\nu}+m^2\alpha_1\left(
\mathcal{K}_{\mu\nu}-tr\mathcal{K}g_{\mu\nu}\right)&+m^2\alpha_2\left[
2\left(tr\mathcal{K}\mathcal{K}_{\mu\nu}-2\mathcal{K}^\rho_\mu\mathcal{K}_{\rho\nu}\right)\right]
\nonumber\\-m^2\alpha_2g_{\mu\nu}\left[\left(tr\mathcal{K}\right)^2-tr\left(\mathcal{K}\right)^2\right]
&=2G\left(F_{\mu\rho}{F_\nu}^\rho-\frac{1}{4}g_{\mu\nu}F^2\right),
\label{eq:gravity}
\end{align}
\begin{align}
\nabla_\mu F^{\mu\nu}=0.
\label{eq:em field}
\end{align}

For solving the equations of motion easily, one can assume that the metric and gauge field are only spherically symmetric, and that the gauge field only contains one scalar potential, $A_\mu=(A_t,0,0,0)$. In addition, the metric can be assumed to be
\begin{align}
\mathrm{d}s^2=-f\left(r\right)\mathrm{d}t^2+\frac{1}{f\left(r\right)}\mathrm{d}r^2+r^2 \mathrm{d}\Omega_2^2.
\end{align}
Putting the assumptions into eqs. ($\ref{eq:gravity}$) and ($\ref{eq:em field}$), one can find the following solutions,
\begin{equation}
f\left(r\right)=1-\frac{2GM}{r}+\frac{GQ^2}{r^2}+\frac{r^2}{l_{\rm AdS}^2}+m^2\alpha_1r+2m^2\alpha_2,
\label{metric function}
\end{equation}

\begin{equation}
A_t=-\frac{Q}{r},\label{Q}
\end{equation}
where $M$ and $Q$ are the mass and total charge of black holes, respectively. In the following we focus only on neutral black holes, so the charge $Q$ is set to be zero.

Based on the above black hole solution, we can construct the shock wave geometry in the conventional way by following Dary and 't Hooft \cite{hooft}. At first, the metric is written in the Kruskal lightcone coordinates in $d$ dimensions,
\begin{align}
\mathrm{d}s^2=-2A\left(u,v\right)\mathrm{d}u\mathrm{d}v+
B\left(u,v\right)\mathrm{d}\Omega_{d-2}^2,
\label{metric:Kruskal}
\end{align}
where $A(u,v)$ and $B(u,v)$ are defined as
\begin{equation}
A(u,v)\equiv -\frac{4}{uv}\frac{f(r)}{[f'\left(r_{\rm h}\right)]^2}, \qquad B(u,v)\equiv r^2.\label{AandB}
\end{equation}
The relationship between the Kruskal coordinates and the spherical coordinates is given by
\begin{align}
u=e^{\frac{2\pi}{\beta}\left[r_*(r)-t\right]},\qquad
v=-e^{\frac{2\pi}{\beta}\left[r_*(r)+t\right]},\nonumber\\
uv=-e^{\frac{4\pi}{\beta}r_*(r)},\qquad u/v=-e^{-\frac{4\pi}{\beta}t},
\label{coord trans}
\end{align}
where $r_*(r)$ is the tortoise coordinate defined as $r_*(r)\equiv \int\frac{\mathrm{d}r}{f(r)}$, and $\beta$ is the inverse temperature of the black hole.
Then, the shock wave geometry can be introduced in such a way that, for $u<0$, the metric eq.~($\ref{metric:Kruskal}$) is kept unchanged;  but for $u>0$, $v$ is replaced by $v+h\left(x^i\right)$ in eq.~($\ref{metric:Kruskal}$),
\begin{align}
\mathrm{d}{\tilde s}^2=-2A\left(u,v+\theta(u)h(x^i)\right)\mathrm{d}u\left(\mathrm{d}v+\theta(u)h,_i\mathrm{d}x^i\right)
+B\left(u,v+\theta(u)h(x^i)\right)\mathrm{d}\Omega_{d-2}^2,
\label{metric of shock1}
\end{align}
where $\theta(u)$ is the Heaviside step function and $h\left(x^i\right)$, sometimes called the shift function, represents a boundary disturbance (perturbation) that only depends on $d-2$ transverse coordinates $x^i$, $i=1,2,\cdots, d-2$. Introducing the following transformation of coordinates,
\begin{align}
u^{\prime}&=u, \nonumber\\
v^{\prime}&=v+\theta\left(u\right)h(x^i),\nonumber\\
x^{{\prime}i}&=x^i,
\end{align}
and substituting it into eq.~($\ref{metric of shock1}$), we obtain the metric in the new coordinates,
\begin{align}
\mathrm{d}{\tilde s}^2=-2A\left(u',v'\right)\mathrm{d}u'\left(\mathrm{d}v'-\delta\left(u'\right)\mathrm{d}u'\right)
+B\left(u',v'\right)\mathrm{d}\Omega_{d-2}^2,
\label{metric of shock2}
\end{align}
where $\delta(u)$ is the Dirac $\delta$-function, and the metrics, eq.~($\ref{metric of shock1}$) and eq.~($\ref{metric of shock2}$), are shown \cite{hooft} to be continuous. Because $A(u,v)$ and $B(u,v)$ are functions of $r_*(r)$ that is function of $uv$ (see eq. (\ref{coord trans})), we thus have the condition,
\begin{align}
\left.\frac{\partial A\left(u,v\right)}{\partial v}\right|_{u=0}=
\left.\frac{\partial B\left(u,v\right)}{\partial v}\right\arrowvert_{u=0}.
\label{condition}
\end{align}

It is convenient to calculate the Ricci tensor $R_{\mu'\nu'}$ by using the metric eq.~($\ref{metric of shock2}$) in the $4$-dimensional spacetime. Then, we transform $R_{\mu'\nu'}$ to its form in the original coordinates $(u,v,\theta,\phi)$, i.e. $R_{\mu\nu}$.  Using the condition eq.~($\ref{condition}$), we can simplify $R_{\mu\nu}$ through an algebraic computation,
\begin{align}
R_{uu}&=\frac{A}{B}\delta(u)\triangle h(\theta,\phi)+\left[\frac{A,_{uv}}{A}
-2\left(\frac{A,_{uv}}{A}+\frac{B,_{uv}}{B}\right)\right]\delta(u)h(\theta,\phi) \nonumber\\
      &+\frac{1}{2B^2}\left(\frac{2A,_uBB,_u}{A}+B,_u^2-2BB,_{uu}\right)\delta(u)h(\theta,\phi),\nonumber\\
R_{uv}&=-\left(\frac{A,_{uv}}{A}+\frac{B,_{uv}}{B}\right),\nonumber\\
R_{u\theta}&=-\theta(u)h,_\theta(\theta,\phi)\left(\frac{A,_{uv}}{A}+\frac{B,_{uv}}{B}\right),\nonumber\\
R_{u\phi}&=-\theta(u)h,_\phi(\theta,\phi)\left(\frac{A,_{uv}}{A}+\frac{B,_{uv}}{B}\right),\nonumber\\
R_{\theta\theta}&=1+\frac{B,_{uv}}{A},\nonumber\\
R_{\phi\phi}&=1+\frac{B,_{uv}}{A}.
\end{align}
Substituting the above components of the Ricci tensor $R_{\mu\nu}$ into the equations of motion eq.~($\ref{eq:gravity}$), we finally derive\footnote{As mentioned under eq.~($\ref{Q}$), we only deal with neutral black holes, so that the right hand side of eq.~($\ref{eq:gravity}$) equals zero. That is, the equations of motion eq.~($\ref{eq:gravity}$) can be simplified to be $E_{\mu\nu}=0$, where the tensor $E_{\mu\nu}$ is defined as the left hand side of eq.~($\ref{eq:gravity}$).}
\begin{align}
E_{uu}&=E^0_{uu}-\frac{A}{B}\delta(u)\triangle h(\theta,\phi)+
         \left[\frac{A,_{uv}}{A}-2\left(\frac{A,_{uv}}{A}
         +\frac{B,_{uv}}{B}\right)\right]\delta(u)h(\theta,\phi),\nonumber\\
E_{u\theta}&=-\theta(u)h,_\theta(\theta,\phi)\left(\frac{A,_{uv}}{A}
         +\frac{B,_{uv}}{B}\right),\nonumber\\
E_{u\phi}&=-\theta(u)h,_\phi(\theta,\phi)\left(\frac{A,_{uv}}{A}
         +\frac{B,_{uv}}{B}\right),\nonumber\\
E_{uv}&=E^0_{uv},\qquad E_{\theta\theta}=E^0_{\theta\theta},\qquad E_{\phi\phi}=E^0_{\phi\phi},
\end{align}
where $E^0_{\mu\nu}$ stands for the tensor without a shock wave and can be omitted directly as it satisfies the vacuum field equation. The stress-energy tensor for a particle located at $u=0$ has only $uu$ component, so that we get $E_{u\theta}=E_{u\phi}=0$ which implies
\begin{align}
\frac{A,_{uv}}{A}+\frac{B,_{uv}}{B}=0.
\end{align}
As a result, only the $uu$ component is non-trivial,
\begin{align}
-\frac{A(0)}{B(0)}\triangle h(\theta,\phi)-\frac{B,_{uv}(0)}{B(0)}h(\theta,\phi)
=8\pi GT_{uu}.
\label{eq:shock}
\end{align}

Let us consider a special case. When the shift function is independent of the transverse coordinates $(\theta, \phi)$, i.e. for the global shock wave, the first term in eq.~($\ref{eq:shock}$) vanishes, we thus obtain the transverse-coordinate-independent shift from eq.~($\ref{eq:shock}$),
\begin{align}
h=\frac{8\pi G A(0)}{A,_{uv}(0)}T_{uu}\propto e^{\frac{2\pi}{\beta}\left(|t_w|-t_*\right)},
\end{align}
where the stress-energy tensor takes the form $T_{uu}=\frac{\mathcal E}{l^2_{\rm AdS}}e^{\frac{2\pi}{\beta}|t_w|}$, and $t_*$ is the scrambling time $t_*=\frac{\beta}{2\pi}\mathrm{ln}\frac{l^2_{\rm AdS}}{cG}$ with $c$ an undetermined coefficient used to absorb the constant in front of the exponent. The stress-energy tensor can be obtained by boosting a particle from $t_w=0$ to $t_w\rightarrow -\infty$ that is located at the position close to the past horizon and has the asymptotic energy $\mathcal E$, where $\mathcal E$ tends to zero while keeping a finite ${\mathcal E}e^{\frac{2\pi}{\beta}|t_w|}$.

\section{Action growth}
\label{sec:growth}
\subsection{The case with no shock waves}
The complexity of a boundary state corresponds to the classical action in the WDW patch. As discussed by Brown et al.~\cite{1512.04993}, the action growth outside a black hole is infinite but independent of time due to the time-translation symmetry of static solution. So the action growth rate\footnote{It is defined as the derivative of action growth with respect to time.} outside the black hole is vanishing, that is, the contribution of this region to the action growth can be omitted. Thus, we just consider the contribution that comes from the regions behind horizons. At the late time, i.e. $t_{\rm L}+t_{\rm R}\gg \beta$, where $t_{\rm L}$ and $t_{\rm R}$ are set to be positive and denote the left and right boundary times, respectively, the contribution from the region behind the past horizon shrinks exponentially to zero in the case of black holes without a shock wave. Therefore, we only need to consider the contribution from the region behind the future horizon.
\begin{figure}
  [!ht]
  \centering
  \includegraphics[totalheight=35mm]{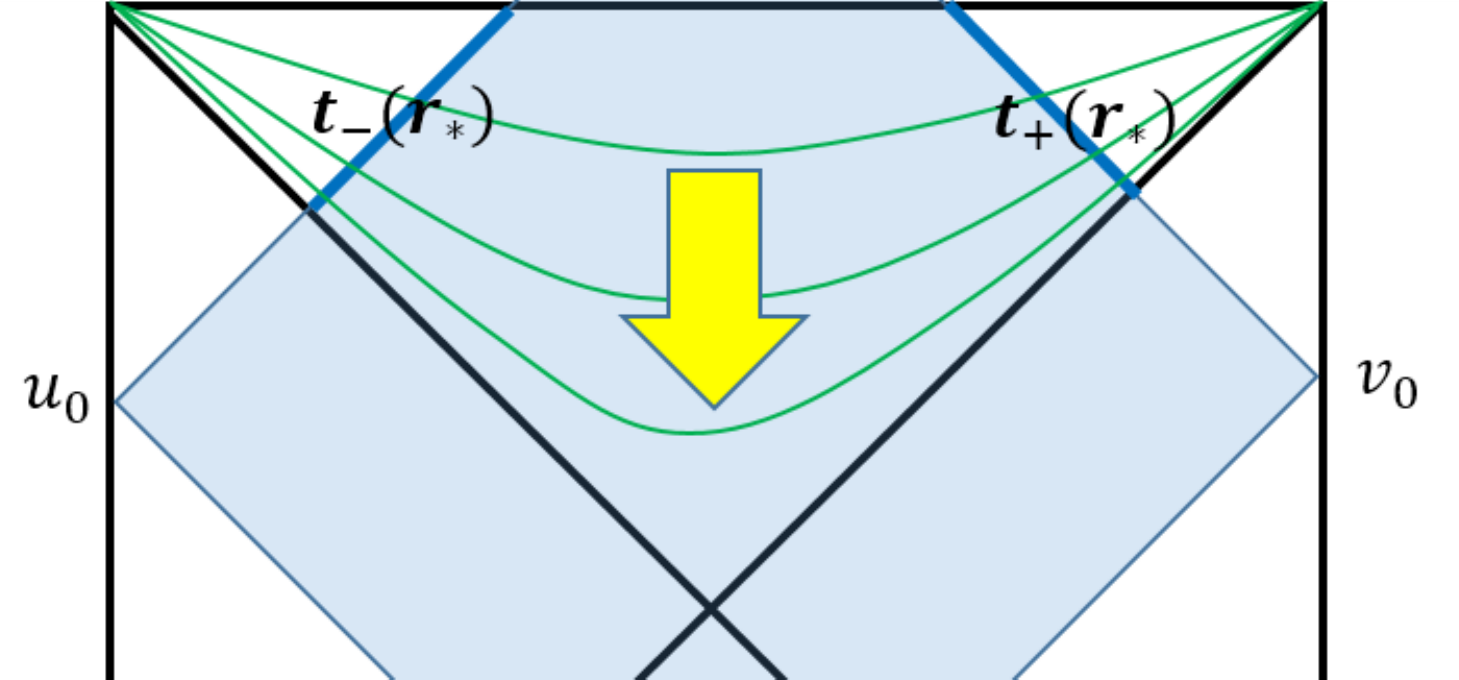}
  \caption{{\footnotesize $t_+(r_*)$ and $t_-(r_*)$ can be determined by the correlations $t_+(r_*)=-r_*+\mathrm{ln}v_0$ and $t_-(r_*)=r_*-\mathrm{ln}u_0$ on a constant $r$ surface.}}
\end{figure}

The gravitational action of interior region is
\begin{align}
A_{\rm bulk}=\Omega_2\int^{r_h}_{0}\int^{t_+(r_*)}_{t_-(r_*)}\mathcal{L}\left(g_{\mu\nu}\right)r^2
\mathrm{d} t\mathrm{d}r,
\end{align}
where $t_-(r_*)$ and $t_+(r_*)$ are the left and right boundaries, respectively, of the WDW patch in a constant $r$ spacelike slice behind the future horizon. See Figure 1 and its caption for the details. Note that the Lagrangian is independent of coordinate $t$ in the case of static solution. So we can work out the $t$ integration directly and keep only the radial integral from the singularity to the black hole radius. To do the time integration, we can re-express the range of integration in terms of $u$ and $v$ by the relation eq.~($\ref{coord trans}$), and then put the metric function eq.~($\ref{metric function}$) into the $r$ integral,
\begin{align}
A_{\rm bulk}&=\Omega_2\frac{\beta}{2\pi}\mathrm{ln}(u_0v_0)
\int^{r_h}_0\mathcal{L}\left(g_{\mu\nu}\right)r^2\mathrm{d}r\nonumber\\
&=-\frac{1}{2G}\left[\frac{r^3_{\rm h}}{l^2_{\rm AdS}}
+\frac{m^2\alpha_1}{2}r^2_{\rm h}\right]\frac{\beta}{2\pi}\mathrm{ln}(u_0v_0),
\label{action:bulk}
\end{align}
where $u_0$ and $v_0$ are left and right boundary values, respectively, and can be written in terms of the boundary times $t_{\rm L}$ and $t_{\rm R}$ as
\begin{equation}
u_0=e^{\frac{2\pi}{\beta}t_{\rm L}}, \qquad v_0=e^{\frac{2\pi}{\beta}t_{\rm R}}.\label{u0v0}
\end{equation}

Similarly, we can derive the boundary contribution by using the trace formula of the extrinsic curvature on a constant $r$ surface, $K=\frac{f'(r)}{2\sqrt{f(r)}}+\frac{2\sqrt{f(r)}}{r}$,
\begin{align}
A_{\rm boundary}=\frac{1}{2G}\left[\frac{r^3_{\rm h}}{l^2_{\rm AdS}}
+\frac{m^2\alpha_1}{2}r^2_{\rm h}+4GM\right]\frac{\beta}{2\pi}\mathrm{ln}(u_0v_0).
\label{action:boundary}
\end{align}

Adding eq.~($\ref{action:bulk}$) and eq.~($\ref{action:boundary}$) together, we obtain the total action growth in the WDW patch,
\begin{align}
A_{\rm WDW}=A_{\rm bulk}+A_{\rm boundary}=2M\frac{\beta}{2\pi}\mathrm{ln}(u_0v_0)=2M\left(t_{\rm L}+t_{\rm R}\right).\label{awdw}
\end{align}
The growth rate of boundary complexity can be calculated by taking the derivative of action with respect to the corresponding boundary time and keeping the other boundary time fixed. For example, we can calculate the action growth rate with respect to the left boundary time as follows,
\begin{align}
\frac{\mathrm{d}A_{\rm WDW}}{\mathrm{d}t_{\rm L}}=2M,
\end{align}
which coincides with the result given by ref.~\cite{1612.03627} in which no shock waves are considered. This result implies that the computational rate of the neutral black hole that saturates the Lloyd bound is the fastest in the nature.

\subsection{The case with a global shock wave}
The initial state of a black hole can be modeled~\cite{0106112} by the thermofield double (TFD) state,  $\left|{\rm TFD}\right\rangle$, and a shock wave sent at time $t_w$ into the bulk spacetime corresponds to the precursor operator $W\left(t\right)$ acting at $t_w$ on the boundary system, where $W\left(t_w\right)$ takes the form, $W\left(t_w\right)=e^{iH_{\rm L}t_w}We^{-iH_{\rm L}t_w}$. With the considerations, the initial state of a black hole can be written as $W\left(t_w\right)\left|{\rm TFD}\right\rangle$, and its time evolution is thus given by
\begin{align}
e^{-iH_{\rm L}t_{\rm L}}e^{-iH_{\rm R}t_{\rm R}}W\left(t_w\right)\left|{\rm TFD}\right\rangle,
\end{align}
where $H_{\rm L}$ and $H_{\rm R}$ are Hamiltonians on the left and right boundaries, respectively,
and $t_w$ tends to $-\infty$.

When the spacetime contains a global shock wave, the Kruskal diagram has two different situations in the late time limit. In one situation, the two boundaries of the WDW patch $\mathcal {M}$ intersect behind the past horizon and in the other situation, the boundaries of the WDW patch $\mathcal {M}$ touch the past singularity. The shape of the WDW patch depends on the value of the shift $h$ which is independent of transverse coordinates. From Figure 1, we can determine the value of a small $h$ by the relation $u^{-1}_0+h<v_0$, and  the value of a large $h$ by the relation $u^{-1}_0+h>v_0$. When considering the late time limit and using eq.~($\ref{u0v0}$), we can rewrite the two inequalities to be $|t_w|-t_*<t_{\rm R}$ and $|t_w|-t_*>t_{\rm R}$. We shall explain at the end of this subsection that the latter relation can be regarded as a criterion to judge whether the right boundary observer can encounter the firewall or not.

As discussed by Susskind \cite{1512.04993}, when the value of $h$ is small,  the action growth emerges merely from the region behind the future horizon,
\begin{align}
A_{|t_w|-t_*\leqslant t_{\rm R}}=2M\left(t_{\rm L}+t_{\rm R}\right),\label{gxh}
\end{align}
because the contribution from the region behind the past horizon tends to zero.  Note that this result is same as that of the case with no shock waves, see eq.~($\ref{awdw}$).

When the value of $h$ is large, both the regions behind the future horizon and the past horizon should be considered. For the region behind the future horizon, the right boundary of the WDW patch is $v_{\rm R}=v_0+h$ due to the existence of the shock wave, so this part of contributions to the action growth reads
\begin{align}
A_{\rm future}=2M\left(t_{\rm L}+|t_w|-t_*\right).
\end{align}
For the region behind the past horizon, the location of the left boundary of the WDW patch is $u_{\rm L}=-v_0^{-1}-h$, and that of the right boundary is $u_{\rm R}=-v_0^{-1}$, where the relation $uv=-1$ at the spacetime boundary has been used. We thus obtain the contribution from the region behind the past horizon,
\begin{align}
A_{\rm past}=2M\left(-t_{\rm R}+|t_w|-t_*\right).
\end{align}
As a result, the total action growth in the WDW patch with a large $h$ is
\begin{align}
A_{|t_w|-t_*\geqslant t_{\rm R}}=A_{\rm future}+A_{\rm past}=2M\left[t_{\rm L}-t_{\rm R}+2\left(|t_w|-t_*\right)\right].\label{ldh}
\end{align}
The first two terms in the bracket correspond the complexity induced by the evolution of the system itself. The complexity grows with respect to the left boundary time $t_{\rm L}$, but decreases with respect to the right boundary time $t_{\rm R}$. This is in agreement with the fact that the shock wave is launched from the left boundary, accesses the black hole, and then approaches the right horizon, which implies that a right boundary observer can meet the shock wave by the time $t_{\rm R}$ under the condition $t_{\rm R} \leqslant |t_w|-t_*$. The terms in the parenthesis stand for the complexity induced by the boundary disturbance. The action growth in $|t_w|$ is twice the growth in $t_{\rm L}$, coming from the fact that $W\left(t_w\right)$ is made up of two time evolution operators, each of which accrues complexity linearly with time. Additionally, the phenomenon that the action growth in $|t_w|$ is delayed by a scrambling time $t_*$ corresponds to the so-called ``switchback" effect \cite{1406.2678}.

\subsection{The case with a local shock wave}
Geometrically, when the spacetime contains a local shock wave, the shift function depends on transverse coordinates,
so the state of the boundary system is given by
\begin{align}
e^{-iH_{\rm L}t_{\rm L}}e^{-iH_{\rm R}t_{\rm R}}W_x(t_w)|{\rm TFD}\rangle,
\end{align}
where $W_x(t_w)$ is the precursor operator, $W_x(t_w)=e^{iH_{\rm L}t_w}W_xe^{-iH_{\rm L}t_w}$, and $W_x$ is localized on the boundary at $x$. Note that for a local shock wave the operator $W_x(t_w)$ grows in spatial directions and the complexity growth due to the boundary disturbance depends on its growth velocity, i.e. the butterfly velocity in spatial directions. 

Because the shift function of a local shock wave depends on transverse coordinates,  we have to solve eq.~($\ref{eq:shock}$). For simplicity but without loss of generality,  we just discuss the $3$-dimensional case in which there exists only one transverse coordinate denoted by $x$. Multiplying eq.~($\ref{eq:shock}$) by the coefficient ${B(0)}/{A(0)}$, we get the desired equation,
\begin{align}
-\frac{\mathrm{d^2}}{\mathrm{d}x^2}h(x)+\mu^2h(x)=
8\pi G\frac{B(0)}{A(0)}e^{\frac{2\pi}{\beta}|t_w|}\delta(x),
\end{align}
where the parameter $\mu^2$ is defined as
\begin{align}
\mu^2\equiv -\frac{B,_{\mu\nu}(0)}{2A(0)}.\label{mu}
\end{align}
The solution can be expressed as
\begin{align}
h(x)\sim e^{\frac{2\pi}{\beta}\left(|t_w|-t_*\right)-\mu |x|}
=e^{\frac{2\pi}{\beta}\left(|t_w|-t_*-\frac{|x|}{v_{\rm B}}\right)},
\label{h:local}
\end{align}
where $v_{\rm B}$ is defined by~\cite{1306.0622,1409.8180}
\begin{align}
v_{\rm B}\equiv \frac{2\pi}{\beta\mu},
\label{butterfly}
\end{align}
which is called the butterfly velocity meaning the spread speed of the local disturbance on the boundary.
Substituting the expression of the shift eq.~($\ref{h:local}$) into the action behind the future horizon,\footnote{The action behind the future horizon takes the form, $A_{\rm future}=2M\frac{\beta}{2\pi}\int\mathrm{ln}(u_0v_{\rm R})\mathrm{d}x$, where $v_{\rm R}$ is the right boundary of the WDW patch, $v_{\rm R}=v_0+h(x)$.} we have
\begin{align}
A_{\rm future}=2M\frac{\beta}{2\pi}\frac{1}{L}\int\mathrm{ln}
e^{\frac{2\pi}{\beta}\left(|t_w|-t_*+t_{\rm L}-\frac{|x|}{v_{\rm B}}\right)}\mathrm{d}x,
\end{align}
where $L\equiv \int\mathrm{d}x$ is the length of the transverse direction that goes to infinite for a planar black hole. Similarly,  substituting the expression of the shift eq.~($\ref{h:local}$) into the action behind the past horizon, we obtain
\begin{align}
A_{\rm past}=2M\frac{\beta}{2\pi}\frac{1}{L}\int\mathrm{ln}
e^{\frac{2\pi}{\beta}\left(|t_w|-t_*-t_{\rm R}-\frac{|x|}{v_{\rm B}}\right)}\mathrm{d}x.
\end{align}
Now we add the above two actions together and choose the upper limit of integral at which the effect of the shock wave tends to zero. Note that the effect of shock waves emerges when the ``large shift condition'' $u_0^{-1}+h(x)\geq v_0$ is guaranteed, that is $|t_w|-t_*-\frac{|x|}{v_{\rm B}}\geq t_{\rm R}$. Thus, the maximal transverse coordinate, $|x|=v_{\rm B}\left(|t_w|-t_*-t_{\rm R}\right)$, should be the upper limit of the integral, and the final result\footnote{In $d$ dimensions the action growth takes the form, $A=2M\left(t_{\rm L}+t_{\rm R}\right)+4{\cal D}v_{\rm B}^{d-2}{\Omega}_{d-3}\frac{\left(|t_w|-t_*-t_{\rm R}\right)^{d-1}}{(d-1)(d-2)}$, where ${\cal D}\equiv M/L^{d-2}$.} is
\begin{align}
A=A_{\rm future}+A_{\rm past}=2M\left(t_{\rm L}+t_{\rm R}\right)+2{\cal D}v_{\rm B}\left(|t_w|-t_*-t_{\rm R}\right)^2,
\label{action:final}
\end{align}
where $\cal{D}$ is the energy density that satisfies ${\cal D}=M/L$ in the transverse direction. As discussed in the beginning of this subsection, we can see from eq.~($\ref{action:final}$) that the action growth depends indeed on the butterfly velocity. In addition, the second term of the action growth depends on $t_{\rm R}$ but not on $t_{\rm L}$ because the local shock wave reaches the right side of black holes, which can be seen geometrically from Figure 1.

Considering the effect of graviton mass terms, we calculate the butterfly velocity as follows. The key point is to work out $\mu$ defined by eq.~($\ref{mu}$).

Because the functions $A(u,v)$ and $B(u,v)$, see eq.~($\ref{AandB}$), are unary functions of variable $uv$, the partial derivative of function $B(u,v)$ with respect to $u$ and $v$ at $uv=0$ can be expressed as $B^\prime(0)$,
\begin{align}
\lim_{u\rightarrow 0}\frac{\partial}{\partial u}\frac{\partial}{\partial v}B(uv)
=\lim_{u\rightarrow 0}\frac{\partial}{\partial u}\left[B^\prime(uv)u\right]
=\lim_{u\rightarrow 0}\left[B^{\prime\prime}(uv)uv+B^\prime(uv)\right]=B^\prime(0).
\end{align}
Thus, $\mu^2$ can be expressed as
\begin{align}
\mu^2=-\frac{B,_{uv}(0)}{2A(0)}=-\frac{B^\prime(0)}{2A(0)}.\label{mu2}
\end{align}
Moreover, considering the conditions,
\begin{equation}
A(0)=-\left.\frac{2}{\kappa}\frac{\mathrm{d}r}{\mathrm{d}(uv)}\right|_{u=0}, \qquad
B^{\prime}(0)=\left.2r_{\rm h}\frac{\mathrm{d}r}{\mathrm{d}(uv)}\right|_{u=0},
\end{equation}
we derive from eq.~(\ref{mu2}) the desired result, $\mu^2={\kappa r_{\rm h}}/{2}$. Substituting this result into the definition of the butterfly velocity eq.~($\ref{butterfly}$), we obtain\footnote{The butterfly velocity was calculated in ref.~\cite{1707.00509} for some 3-dimensional gravity models.}
\begin{align}
v_{\rm B}=\sqrt{\frac{\kappa}{2r_{\rm h}}}.
\end{align}

Now we compute the butterfly velocity for the Einstein gravity and the massive gravity. For the former with the metric function, $f(r)=1-\frac{2GM}{r}+\frac{r^2}{l^2_{\rm AdS}}$, and the surface gravity, $\kappa={f^{\prime}(r_{\rm h})}/{2}$, the butterfly velocity reads
\begin{align}
{\tilde v}_{\rm B}=\sqrt{\frac{1}{2}\left(\frac{1}{l^2_{\rm AdS}}+\frac{GM}{r^3_{\rm h}}\right)}.
\end{align}
For the latter with the metric function, $\hat{f}(r)=1-\frac{2GM}{r}+\frac{r^2}{l_{\rm AdS}^2}+m^2\alpha_1r+2m^2\alpha_2$, where the hat labels quantities corresponding to the massive gravity, the butterfly velocity takes the form,
\begin{align}
\hat{v}_{\rm B}=\sqrt{\frac{1}{2}\left(\frac{1}{l^2_{\rm AdS}}
+\frac{GM}{\hat{r}^3_{\rm h}}+\frac{m^2\alpha_1}{\hat{r}_{\rm h}}\right)},
\end{align}
where $\hat{r}_{\rm h}$ is the solution of equation $\hat{f}\left(\hat{r}_{\rm h}\right)=0$.

Next we compare the butterfly velocities of the two gravity theories and then determine the size relation of the corresponding action growths. Because the both coupling constants $\alpha_1$ and $\alpha_2$ are negative, the metric function in the massive gravity is smaller than that in the Einstein gravity when $M$ and $l_{\rm AdS}$ are fixed, namely,
\begin{align}
\hat{f}(r)-f(r)=m^2\alpha_1r+2m^2\alpha_2<0.
\label{diff}
\end{align}
As the values of horizon radii $r_{\rm h}$ and $\hat{r}_{\rm h}$ are determined by equations $f(r_{\rm h})=0$ and $\hat{f}\left(\hat{r}_{\rm h}\right)=0$, respectively, we can deduce by using eq.~(\ref{diff}),
\begin{align}
\hat{f}(r_{\rm h})<0=\hat{f}\left(\hat{r}_{\rm h}\right).
\label{inequ}
\end{align}
Note that the metric function $\hat{f}(r)$ is negative when $r$ is smaller than $\hat{r}_{\rm h}$, so eq.~(\ref{inequ}) implies
\begin{align}
r_{\rm h} < \hat{r}_{\rm h},
\end{align}
which gives rise to
\begin{align}
\frac{1}{l^2_{\rm AdS}}+\frac{M}{\hat{r}^3_{\rm h}}<\frac{1}{l^2_{\rm AdS}}+\frac{M}{r^3_{\rm h}}.
\end{align}
In addition, considering ${m^2\alpha_1}/{\hat{r}_{\rm h}}<0$,
we can definitely determine
\begin{align}
\hat{v}_{\rm B}<{\tilde v}_{\rm B}.\label{mlesse}
\end{align}
This means that the butterfly velocity in the massive gravity is smaller than that in the Einstein gravity. As a result, the action growth or the complexity in the massive gravity is less than that in the Einstein gravity due to the effect of the graviton mass in accordance with eq.~(\ref{action:final}). Because the action growth rate is defined as the derivative of the action growth with respect to time, the same speculation can be made for the action growth rate in the bulk or the computational rate on the boundary.

\section{Implication of C/A duality from a new calculation method of boundary terms}
Parattu et al. demonstrated~\cite{1212.2922,1501.01053} that the York-Gibbons-Hawking boundary term is a failure when the boundary is null-like. Furthermore, Lehner et al. proposed~\cite{1609.00207} that extra terms are needed when the boundary is non-smooth. The boundary term of the WDW patch is null-like at the future horizon and the intersection of two past boundaries behind the past horizon is singular. Therefore, the following counter terms should be added,
\label{sec:volume}
\begin{align}
\frac{1}{8\pi G}\int_\mathcal{N}\mathrm{d}^3x\sqrt{-g}\lambda
+\frac{1}{8\pi G}\oint_{\mathcal{B}}\mathrm{d}^2x\sqrt{\gamma}a,
\end{align}
which can overcome the problems caused by the null-like boundaries and the singularity.
Here the notations should be explained. $\mathcal{N}$ is a null-like boundary and $\mathcal{B}$ is the intersection of two past boundaries. $\gamma_{ab}$ is the reduced 2-metric on $\mathcal{B}$. The definition of $a$ is $a\equiv \mathrm{ln}\left(\bf{k}\cdot\bf{\bar{k}}\right)$, where $\bf{k}$ and $\bf{\bar{k}}$ are the null normals to the corner pieces defined as $k_a\equiv -c\nabla_av$ and $\bar{k}_a\equiv \bar{c}\nabla_au$ with constants $c$ and $\bar{c}$, and $\lambda$ is defined by
\begin{equation}
k^a\nabla_ak^b\equiv \lambda k^b,
\end{equation}
or
\begin{equation}
\lambda\equiv -n_bk^a\nabla_ak^b,
\end{equation}
when an auxiliary vector $n_a$ satisfying $\bf{n}\cdot \bf{k}=-\mathrm{1}$ is introduced.

Following refs.~\cite{1212.2922,1501.01053}, Lehner et al. gave~\cite{1609.00207}  a detailed argument that the null boundaries lying on the horizon do not contribute to the action growth. But in Brown's calculation~\cite{1512.04993} the null boundaries have their contributions, where the WDW patch is divided into two pieces and only the one behind the horizon is considered. This seeming contradiction can be explained clearly. As the future horizon divides a region into two sub-regions, it can be regarded as the common boundary of the two sub-regions. Actually, the boundary terms of the two sub-regions cancel each other. In addition, since the boundary of the WDW patch is not smooth, one has to take into account the contribution located at the corner (the blue point in Figure 2). As a summary, the action should be calculated by considering the contributions from the interior of the WDW patch, the boundary located at the future singularity which is spacelike, and the corner of the past boundary of the WDW patch. The corresponding regions are depicted in Figure 2. Couch et al. recalculated~\cite{1610.02038} the action in the Einstein gravity and found that all terms can be expressed as thermodynamical quantities. They also presented a new conjecture called ``C/V duality 2.0''. In the Einstein gravity, the action growth comes from the regions behind the horizon, the boundary located in the future singularity, and the corner of the past boundary. The contributions can be expressed as $-PV$, ${3M}/{2}$, and $TS$, respectively. The total contribution of the three terms equals ${3M}/{2}+TS-PV=2M$, which is in fact the Smarr formula $M=2TS-2PV$. This result coincides with that obtained by Brown et al. \cite{1512.04993}.
\begin{figure}
  \centering
  \includegraphics[totalheight=55mm]{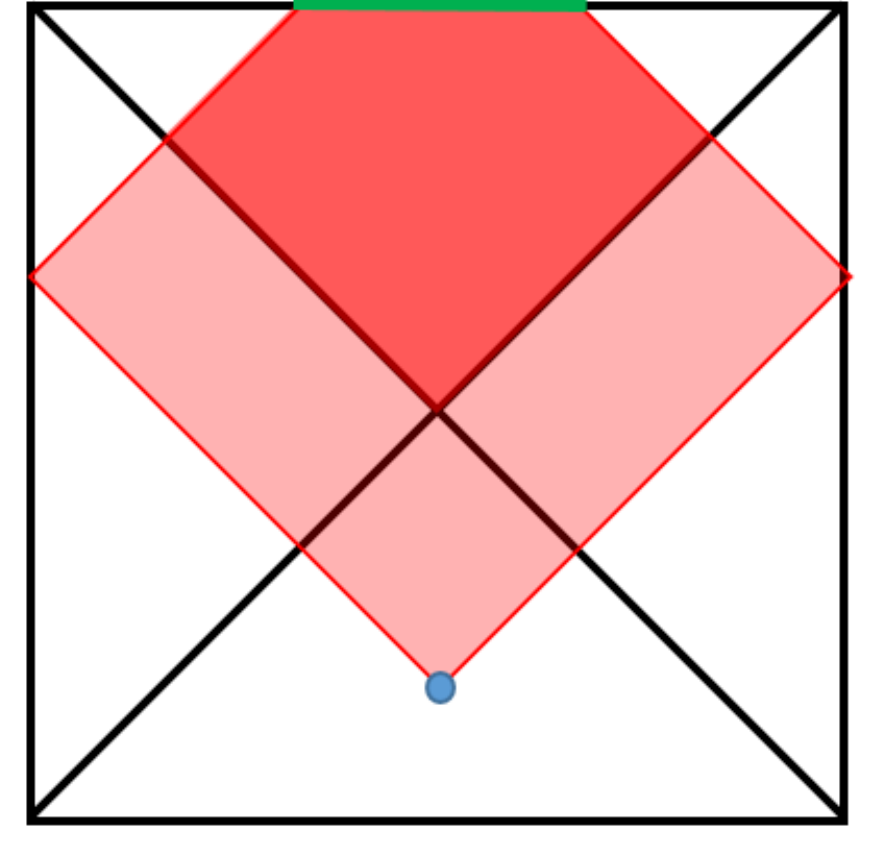}
  \caption{{\footnotesize The interior contribution (dark red region), the spacelike boundary contribution (green line), and the boundary singularity contribution (blue point) of action growth are shown in this figure.}}
\end{figure}

Now we extend the above result to the massive gravity theory. Using the result eq.~($\ref{action:bulk}$) directly, we can write the bulk contribution in the form,
\begin{align}
A_{\rm bulk}=-\left[PV+\frac{1}{2}\left(\frac{\partial M}{\partial \alpha_1}\right)_{S,P}\alpha_1\right]\left(t_{\rm L}+t_{\rm R}\right),\label{baction}
\end{align}
where $P={3}/({8\pi Gl_{\rm AdS}^2})$ and $V={4}\pi r_{\rm h}^3/{3}$ as usual. If we regard $m^2\alpha_1\equiv Z$ as a new thermodynamic quantity, its conjugate can be defined as $\left(\frac{\partial M}{\partial Z}\right)_{S,P}\equiv Y$. Then the bulk action can be rewritten manifestly in terms of thermodynamic quantities,
\begin{equation}
A_{\rm bulk}=-\left(PV+\frac{1}{2}YZ\right)\left(t_{\rm L}+t_{\rm R}\right).
\end{equation}
As a result, we extend the first law of thermodynamics to be
\begin{align}
\mathrm{d}M=T\mathrm{d}S+V\mathrm{d}P+Y\mathrm{d}Z,
\label{first law}
\end{align}
where the dimensional scaling of $Z$ is the inverse of length. Note that $\alpha_2$ cannot be regarded as a thermodynamic quantity because it does no appear in the bulk action, or in other words, it does not appear in the boundary complexity from the point of view of holographic duality.

The boundary contribution is
\begin{align}
A_{\rm boundary}=-\frac{1}{2G}
\left.\left[\frac{r^2f'(r)}{2}+2r^2f(r)\right]\right|_{r=0}\left(t_{\rm L}+t_{\rm R}\right)
=\frac{3M}{2}\left(t_{\rm L}+t_{\rm R}\right),
\end{align}
which comes only from the future spacelike singularity. In addition, the contribution from the corner reads
\begin{align}
2\oint_{{\mathcal{B}}^{\prime}}\mathrm{d}^2x\sqrt{\gamma}a-2\oint_{\mathcal B}\mathrm{d}^2x\sqrt{\gamma}a
=2\Omega_2\left.\left[r^2\mathrm{ln}\left(-\frac{f}{c\bar{c}}\right)\right]\right|^{r_{{{\mathcal{B}}^{\prime}}}}_{r_{\mathcal B}},
\end{align}
where ${\mathcal B}$ and ${{\mathcal{B}}^{\prime}}$ are intersections corresponding to the past boundary attached to the left boundary at time $t_{\rm L}$ and at time $t_{\rm L}+\delta t$, respectively. When the left boundary time $t_{\rm L}$ changes, i.e. the left intersection between the WDW patch and the boundary of spacetime (see Figure 2) moves up or down, the location of the corner moves along the direction $v$. This means that $\mathrm{d}u=0$, $\mathrm{d}v=\delta t$, and $\mathrm{d}r=-\frac{1}{2}f\delta t$. Therefore, the corner contribution can be reduced to be
\begin{align}
2\Omega_2\left.\left[r^2\mathrm{ln}\left(-\frac{f}{c\bar{c}}\right)\right]\right|^{r_{{\mathcal{B}}^{\prime}}}_{r_{\mathcal B}}
&=-\Omega_2f\left.\frac{\mathrm{d}f}{\mathrm{d}r}\right|_{r=r_{\mathcal B}}\delta t\nonumber\\
&=\Omega_2\left.\left[r^2\frac{\mathrm{d}f}{\mathrm{d}r}+
 2rf\mathrm{ln}\left(-\frac{f}{c\bar{c}}\right)\right]\right|_{r=r_{\mathcal B}}\delta t.
\end{align}
Note that we focus on the case of the late time limit, where the corner is close to the bifurcate. In this limit, the second term in the bracket tends to zero and the first term to $TS$ generally when the black hole has the spherical symmetry. Combining the above results together, we obtain ${3M}/{2}+TS-PV-\frac{1}{2}YZ=2M$. It is in fact the Smarr formula,
\begin{align}
M=2TS-2PV-YZ,
\label{smarr}
\end{align}
which can also be derived by using the scaling law of homogeneous functions directly from the extended first law eq.~($\ref{first law}$). Note that the dimensional scaling of $m^2\alpha_2$ is zero, so $\alpha_2$ does not appear in the Smarr formula, which coincides with the above mentioned observation that $\alpha_2$ does not appear in the bulk action eq.~($\ref{baction}$).

Consequently, we regard $Y$ and $Z$ as a new pair of thermodynamic quantities conjugate to each other and give the extended first law of thermodynamics by adopting the C/A duality and admitting~\cite{1610.02038} the point of view that the quantity, i.e. $YZ$ that appears in holographic complexity should be a thermodynamic quantity.

\section{Conclusion}
\label{sec:conclusion}
We investigate the C/A duality of shock wave geometry in the massive gravity theory, including the global and local shock waves.

In the case of a global (spherically symmetric) shock wave, the action growth is the same for both the Einstein gravity and the massive gravity.  It contains the Hamiltonian evolutions of the boundaries and of the global shock wave, see eqs.~($\ref{gxh}$) and ($\ref{ldh}$) that correspond to a small and large transverse-coordinate-independent shift $h$, respectively. The contribution of the boundaries is proportional to the left and right boundary times, $t_{\rm L}$ and $t_{\rm R}$. For a small shift $h$, the action growth is same as that of the case with no shock waves, that is, the global shock wave has no contribution to the action growth. For a large shift $h$, the action growth in $|t_w|-t_*$ is twice that in $t_{\rm L}-t_{\rm R}$, that is, the global shock wave has the contribution twice that of the boundary. We can think that this originates from the double evolutional time of the precursor operator $W\left(t_w\right)=e^{iH_{\rm L}t_w}We^{-iH_{\rm L}t_w}$. Incidentally, our result characterized by the twice deduction by the scrambling time agrees with the quantum circuit model~\cite{1402.5674}.
In addition, when the spacetime contains several global shock waves, the action growth is proportional to the fold-time $t_f$ \cite{1201.3664,1512.04993} subtracted by the double scrambling time for each shock wave, which coincides with the analysis of the boundary theory.

In the case of a local shock wave, the action growth takes the same form for both the Einstein gravity and the massive gravity, see eq.~($\ref{action:final}$) whose second term is proportional to the butterfly velocity $v_{\rm B}$. Since the boundary disturbance depends on the transverse coordinate in this case, i.e. the shift $h(x)$ is a function of the transverse coordinate $x$, the graviton mass leads to the effect that the butterfly velocity in the massive gravity is smaller than that in the Einstein gravity, see eq.~($\ref{mlesse}$) and its preceding analysis. We thus conclude that the action growth or the complexity in the massive gravity is less than that in the Einstein gravity. In other words, the action growth or the complexity in the massive gravity is depressed by the graviton mass, so is the action growth rate or the computational rate.

Finally, we recalculate the action growth by using the method proposed by Lehner et al. \cite{1609.00207} and express it in terms of thermodynamic quantities as done by Couth et al.~\cite{1610.02038}. Admitting the point of view that the quantity that appears in holographic complexity should be a thermodynamic quantity, we generalize the first law of the massive Schwarzschild-AdS black hole, see eq.~($\ref{first law}$). Moreover, we give the Smarr formula by reconciling the method by Lehner et al. \cite{1609.00207} and that by Couth et al.~\cite{1610.02038}, which further supports the C/A duality.

\section*{Acknowledgments}
Y-GM would like to thank P. Nicolini of Frankfurt Institute for Advanced Studies (FIAS) for kind hospitality. This work was supported in part by the National Natural Science Foundation of China under grant No.11675081. Finally, the authors would like to thank the anonymous referee for the helpful comments that indeed greatly improve this paper.


\begin{thebibliography}{10}
    \bibitem{9711200} J. Maldacena, {\it {The large N limit of superconformal field theories and supergravity}}, Adv. Theor. Math. Phys. {\bf {2}} (1998) 231
        [\href{http://arxiv.org/abs/hep-th/9711200} {\tt{arXiv:hep-th/9711200}}].
    \bibitem{1306.0622} S.H. Shenker and D. Stanford, {\it {Black holes and the butterfly effect}}, JHEP {\bf {03}} (2014) 067
        [\href{http://arxiv.org/abs/1306.0622}{{\tt arXiv:1306.0622[hep-th]}}].
    \bibitem{1403.5695} L. Susskind, {\it{Addendum to computational complexity and black hole horizons}}, Fortschr. Phys. {\bf {64}} (2016)  44 [\href{http://arxiv.org/abs/1403.5695}{{\tt arXiv:1403.5695[hep-th]}}].
    \bibitem{1507.02287} L. Susskind, {\it{The typical-state paradox: Diagnosing horizons with complexity}}, Fortschr. Phys. {\bf {64}}  (2016) 84 [\href{http://arxiv.org/abs/1507.02287}{{\tt arXiv:1507.02287[hep-th]}}].
    \bibitem{1406.2678} D. Stanford and L. Susskind, {\it{Complexity and shock wave geometries}}, Phys. Rev. {\bf {D 90}} (2014) 126007  [\href{http://arxiv.org/abs/1406.2678}{{\tt arXiv:1406.2678[hep-th]}}].
    \bibitem{1509.07876} A. Brown, D.A. Roberts, L. Susskind, B. Swingle, and Y. Zhao, {\it{Holographic complexity equals bulk action?}}  Phys. Rev. Lett. {\bf {116}} (2016) 191301 [\href{http://arxiv.org/abs/1509.07876}{{\tt arXiv:1509.07876[hep-th]}}].
    \bibitem{1512.04993} A. Brown, D.A. Roberts, L. Susskind, B. Swingle, and Y. Zhao, {\it{Complexity, action, and black holes}}, Phys. Rev. {\bf {D 93}} (2016) 086006 [\href{http://arxiv.org/abs/1512.04993}{{\tt arXiv:1512.04993[hep-th]}}].
    \bibitem{Lloyd} S. Lloyd, {\it{Ultimate physical limits to computation}}, Nature
         {\bf {406}}  (2000) 1047.
    \bibitem{1606.08307} R.-G. Cai, S.-M. Ruan, S.-J. Wang, R.-Q. Yang,
        and R.-H. Peng, {\it{Action growth for AdS black holes}}, JHEP {\bf {09}} (2016) 161
        [\href{http://arxiv.org/abs/1606.08307}{{\tt arXiv:1606.08307[gr-qc]}}].
    \bibitem{1610.08063} S. Chapman, H. Marrochio, and R.C. Myers, {\it{Complexity of formation in holography}}, JHEP {\bf {01}} (2017) 062
        [\href{https://arxiv.org/abs/1610.08063}{{\tt arXiv:1610.08063[hep-th]}}].
    \bibitem{1612.00433} D. Carmi, R.C. Myers, and P. Rath, {\it{Comments on holographic complexity}}, JHEP {\bf {03}} (2017) 118
        [\href{https://arxiv.org/abs/1612.00433}{{\tt arXiv:1612.00433[hep-th]}}].
    \bibitem{1702.06766} R.-G. Cai, M. Sasaki, and S.-J. Wang, {\it{Action growth of charged black holes with a single horizon}}, Phys. Rev. {\bf {D 95}} (2017) 124002
        [\href{https://arxiv.org/abs/1702.06766}{{\tt arXiv:1702.06766[gr-qc]}}].
    \bibitem{1702.06796} M. Alishahiha, A.F. Astaneh, A. Naseh, and M.H. Vahidinia, {\it{On complexity for higher derivative gravities}}, JHEP {\bf {05}} (2017) 009
        [\href{https://arxiv.org/abs/1702.06796}{{\tt arXiv:1702.06796[hep-th]}}].
    \bibitem{1703.06297} J. Tao, P. Wang, and H.-T. Yang, {\it{Testing holographic conjectures of complexity with Born-Infeld black holes}},
        \href{https://arxiv.org/abs/1703.06297}{{\tt arXiv:1703.06297[hep-th]}}.
    \bibitem{1703.10468} W.-D. Guo, S.-W. Wei, Y.-Y. Li, and Y.-X. Liu, {\it{Complexity growth rates for AdS black holes in massive gravity and f(R) gravity}},
        \href{https://arxiv.org/abs/1703.10468}{{\tt arXiv:1703.10468[gr-qc]}}.
    \bibitem{1602.08272} E. Perlmutter, {\it{Bounding the space of holographic CFTs with chaos}}, JHEP {\bf {10}} (2016) 069
        [\href{https://arxiv.org/abs/1602.08272}{{\tt arXiv:1602.08272[hep-th]}}].
    \bibitem{1610.02890} M. Alishahiha, A. Davody, A. Naseh, and S.F. Taghavi, {\it{On butterfly effect in higher derivative gravities}}, JHEP {\bf {11}} (2016) 032
        [\href{https://arxiv.org/abs/1610.02890}{{\tt arXiv:1610.02890[hep-th]}}].
    \bibitem{1011.1232} C. de Rham, G. Gabadadze, and A.J. Tolley, {\it{Resummation of massive gravity}}, Phys. Rev. Lett. {\bf {106}}  (2011) 231101
        [\href{http://arxiv.org/abs/1011.1232}{{\tt arXiv:1011.1232[hep-th]}}].
    \bibitem{1610.02038} J. Couch, W. Fischler, and P.H. Nguyen, {\it{Noether charge, black hole volume, and complexity}}, JHEP {\bf {03}} (2017) 119
        [\href{http://arxiv.org/abs/1610.02038}{{\tt arXiv:1610.02038[hep-th]}}].
    \bibitem{1301.0537} D. Vegh, {\it{Holography without translational symmetry}}, \href{http://arxiv.org/abs/1301.0537}{{\tt arXiv:1301.0537[hep-th]}}.
    \bibitem{1409.2369} R.-G. Cai, Y.-P. Hu, Q.-Y. Pan, and Y.-L. Zhang, {\it{Thermodynamics of black holes in massive gravity}}, Phys. Rev. {\bf {D 91}} (2015) 024032
        [\href{http://arxiv.org/abs/1409.2369}{{\tt arXiv:1409.2369[hep-th]}}].
    \bibitem{1511.04967} L.-M. Cao, Y. Peng, and Y.-L. Zhang, {\it{de Rham-Gabadadze-Tolley massive gravity with degenerate reference metrics}}, Phys. Rev. {\bf {D 93}} (2016) 124015
        [\href{http://arxiv.org/abs/1511.04967}{{\tt arXiv:1511.04967[hep-th]}}].
    \bibitem{hooft} T. Dray and G. 't Hooft, {\it{The gravitational shock wave of a massless particle}}, Nucl. Phys. {\bf {B 253}} (1985) 173.
    \bibitem{1612.03627} W.-J. Pan and Y.-C. Huang, {\it{Holographic complexity and action growth in massive gravities}}, Phys. Rev. {\bf D 95} (2017) 126013
        [\href{http://arxiv.org/abs/1612.03627}{{\tt arXiv:1612.03627[hep-th]}}].
        \bibitem{0106112} J. Maldacena, {\it{Eternal black holes in anti-de Sitter}}, JHEP {\bf {04}} (2003) 021
        [\href{http://arxiv.org/abs/hep-th/0106112}{{\tt arXiv:hep-th/0106112}}].
    \bibitem{1409.8180} D.A. Roberts, D. Stanford, and L. Susskind, {\it{Localized shocks}}, JHEP {\bf {03}} (2015) 051
        [\href{http://arxiv.org/abs/1409.8180}{{\tt arXiv:1409.8180[hep-th]}}].
      \bibitem{1707.00509} M.M. Qaemmaqami, {\it{On the butterfly effect in 3d gravity}},
        \href{http://arxiv.org/abs/1707.00509}{{\tt arXiv:1707.00509[hep-th]}}.
    \bibitem{1212.2922} Y. Neiman, {\it{On-shell actions with lightlike boundary data}},
        \href{http://arxiv.org/abs/1212.2922}{{\tt arXiv:1212.2922[hep-th]}}.
    \bibitem{1501.01053} K. Parattu, S. Chakraborty, B.R. Majhi, and T. Padmanabhan, {\it{A boundary term for the gravitational action with null boundaries}}, Gen. Relativ. Gravit. {\bf {48}} (2016) 94
        [\href{http://arxiv.org/abs/1501.01053}{{\tt arXiv:1501.01053[gr-qc]}}].
    \bibitem{1609.00207} L. Lehner, R.C. Myers, E. Poisson, and R.D. Sorkin, {\it{Gravitational action with null boundaries}}, Phys. Rev. {\bf D {94}} (2016) 084046
        [\href{http://arxiv.org/abs/1609.00207}{{\tt arXiv:1609.00207[hep-th]}}].
  \bibitem{1402.5674} L. Susskind, {\it{Computational complexity and black hole horizons}}, Fortschr. Phys. {\bf {64}} (2016) 24
        [\href{http://arxiv.org/abs/1402.5674}{{\tt arXiv:1402.5674[hep-th]}}].   
    \bibitem{1201.3664} I. Heemskerk, D. Marolf, J. Polchinski, and J. Sully, {\it{Bulk and transhorizon measurements in AdS/CFT}}, JHEP {\bf {10}} (2012) 165
        [\href{http://arxiv.org/abs/1201.3664}{{\tt arXiv:1201.3664[hep-th]}}].
    \end{thebibliography}
\end{document}